\documentclass{aa}  

\usepackage{graphicx,xcolor}
\usepackage{hyperref}
\usepackage{mathtools}
\usepackage{threeparttable}
\usepackage{txfonts}

\def\sgras{{Sgr~A$^\star$\ }}
\def\gravity{{Gravity}}

\begin{document}

   \title{An acceleration-radiation model for nonthermal flares from Sgr~A*}


   \author{Maria Petropoulou
          \inst{1,2} 
          \and
          Gabriele Ponti\inst{3,4}
          \and 
          Giovanni Stel \inst{5,4} 
         \and 
         Apostolos Mastichiadis\inst{1}
          }

   \institute{Department of Physics, National and Kapodistrian University of Athens, University Campus Zografos, GR 15784, Athens, Greece 
   \and 
   Institute of Accelerating Systems \& Applications, University Campus Zografos, GR 15784, Athens, Greece \\
              \email{mpetropo@phys.uoa.gr}
         \and
    Max-Planck-Institut für extraterrestrische Physik, Giessenbachstrasse, 85748, Garching, Germany
        \and
    INAF -- Osservatorio Astronomico di Brera, Via E. Bianchi 46, 23807 Merate, Italy
        \and
    DiSAT, Universit\`a degli Studi dell'Insubria, via Valleggio 11, I-22100 Como, Italy    
             }

   \date{}

 
  \abstract
   {\sgras is the electromagnetic counterpart of the accreting supermassive black hole in the Galactic center. Its emission is variable in the near-infrared (NIR) and X-ray wavelengths on short timescales (several minutes to a few hours). The NIR light curve displays red-noise variability, while the X-ray light curve exhibits bright flares that rise by many orders of magnitude upon the stable X-ray quiescent emission. Every X-ray flare is associated with a bright NIR flux change, but the opposite is not always true. The physical origin of NIR and X-ray flares is still under debate.}
   {We introduce a model for the production of NIR and X-ray flares from an active region in \sgras, where particle acceleration takes place intermittently. A fraction of electrons from their thermal pool is accelerated to higher energies while they radiate via synchrotron and synchrotron self-Compton (SSC) processes. In contrast to other radiation models for \sgras flares, the particle acceleration is not assumed to be instantaneous.}
   {We studied the evolution of the particle distribution and the emitted electromagnetic radiation from the flaring region by numerically solving the kinetic equations for electrons and photons. Our calculations took the finite duration of particle acceleration, radiative energy losses, and physical escape from the flaring region into account. To gain better insight into the relation of the model parameters, we complemented our numerical study with analytical calculations.}
   {Flares are produced when the acceleration episode has a finite duration. The rising part in the light curve of a flare is related to the particle acceleration timescale, while the decay is controlled by the cooling or escape timescale of particles. The emitted synchrotron spectra are power laws whose photon index  is determined by the ratio of the acceleration and escape timescales, followed by an exponential cutoff. This occurs at the characteristic synchrotron photon energy emitted by particles with the maximum Lorentz factor (where energy loss and gain rates become equal). The NIR flux increases before the onset of the X-ray flare, and the time lag is linked to the particle acceleration timescale. Bright X-ray flares, such as the one observed in 2014, have $\gamma$-ray counterparts that might be detected by the Cherenkov Telescope Array Observatory.
   }
   {Our generic model for NIR and X-ray flares favors an interpretation of diffusive nonresonant particle acceleration in magnetized turbulence. If direct acceleration by the reconnection electric field in macroscopic current sheets causes the energization of particles during flares in \sgras, then models considering the injection of preaccelerated particles into a blob where particles cool and/or escape would be appropriate to describe the flare.}
   \keywords{Galaxy: center - infrared: general - radiation mechanisms: non-thermal }

   \maketitle
\section{Introduction}
\sgras, the radiative counterpart of the supermassive black hole at the Galactic center, is considered to be one of the best targets for studying accretion physics at low luminosity \citep{Genzel2010}.
The combination of its low bolometric luminosity (about $L\sim10^{36}$ erg s$^{-1}$) and close distance to us makes it one of the few low-luminosity active galactic nuclei (AGN) on which we can study accretion physics at exceptionally low Eddington ratios ($\sim10^{-8}$) in great detail. 

The electromagnetic spectrum of \sgras peaks in the submillimeter band, where it forms the so-called submm bump \citep{Yuan2003}, which has recently been imaged by the Event Horizon Telescope Collaboration \citep{EHT_2022}. It is thought to be due to optically thick synchrotron radiation produced by relativistic  quasi-thermal electrons with a temperature 
$T_e\sim$ few $10^{10}$~K (or typical Lorentz factor $\gamma \sim10$) and a density $n_e\sim10^6$ cm$^{-3}$, embedded in a magnetic field with a strength of $\sim10 - 50$~G \citep{Loeb2007, Genzel2010}.

In the near-infrared (NIR) band, \sgras displays a red-noise light curve at frequencies higher than a fraction of a day, showing continuous variations with multiple peaks and no obvious quiescent state \citep{Do_2009, Witzel_2012, Witzel_2018}. With the \gravity\ interferometer, it has been possible to determine the evolution of the astrometry and polarization of the source during bright NIR  events \citep{gravity2018flares, GRAVITY2020}. During flares, the NIR source appears to be compact ($R<2-3~R_S$)\footnote{The Schwarzschild radius of \sgras is $R_{S}=2 GM/c^2 \simeq 1.2 \times 10^{12}$~cm, for a black hole mass of $4.3\times 10^6 M_\odot$~\citep{GRAVITY2019}.} and to move at $\sim30$ \% of the speed of light. It traces loops on the sky with a radius of about six to ten gravitational radii. At the same time, the NIR polarization angle is observed to rotate. The swings in the polarization angle are consistent with a model in which the synchrotron-emitting flaring region is embedded in a poloidal magnetic field configuration and rotates approximately consistently with the observed astrometric loops \citep{gravity2018flares, GRAVITY2020}.  

In the X-ray band, \sgras displays a quiescent luminosity of $L_{2-10~\rm keV} \sim 2 \times 10^{33}$ erg s$^{-1}$ \citep{Baganoff_2003, Xu_2006}. This quiescent emission is thought to be produced via bremsstrahlung radiation from a hot plasma with a density of $n_e\sim10^{-3}$~cm$^{-3}$ and a temperature $T\sim7\times10^7$~K in a region with a size of $\sim10^5~R_S$. It is likely associated with the accretion of winds from nearby massive stars \citep{Melia_1992,Quataert_2002,Cuadra_2006,Cuadra_2008}. Upon this stable quiescent emission, the X-ray light curve exhibits bright flares that typically last for a small fraction of the time (from several minutes to a few hours). This suggests that flares are individual events that randomly punctuate an otherwise quiescent source \citep{Neilsen_2013,Ponti_2015}. Moreover, it is observed that every X-ray flare has an NIR counterpart (i.e., it is associated with a bright NIR flux excursion), but most NIR flares have no X-ray counterpart. 

X-ray flares have a power-law luminosity distribution (${\rm d}N/{\rm d}L$) with a slope of $\sim -1.9$, and the moderate flares ($L_{2-8 \,  \rm keV} \approx 10^{34}$ erg s$^{-1}$) occur about once per day \citep{Degenaar2013, Neilsen_2013, Neilsen_2015, Ponti_2015}.  
It is still debated whether the flaring rate is stationary or changes over time, or if the flares preferentially occur in clusters \citep{Ponti_2015, Yuan_2016, Mossoux_2016, Mossoux_2017,Bouffard_2019, Andres_2022}. The duration and fluence of the X-ray flares have power-law distributions with observed fluences in the range $\sim10^{-9}-10^{-7}$~erg~cm$^{-2}$ and durations from several minutes to a few hours \citep{Neilsen_2013, Ponti_2015, Yuan_2016}.

According to the canonical szenario for the production of flares at NIR and X-ray energies, a small region in the accretion flow appears to energize electrons well beyond their typical thermal energy. While synchrotron radiation of nonthermal electrons causes the observed NIR emission \citep{Genzel2010}, the origin of the X-ray emission is still debated. Several radiative mechanisms have been proposed to explain the X-ray flaring emission, including synchrotron, synchrotron
self-Compton (SSC), and inverse Compton (IC) on submm seed photons within tens of Schwarzschild radii from \sgras \citep{Markoff_2001, Yuan2003, Eckart_2004, Eckart_2008, Yusef-Zadeh_2006,Yusef-Zadeh_2008, Hornstein_2007, Marrone_2008,  Dodds-Eden_2010,Trap_2011, Dibi_2014,  Gutierrez_2020, Witzel_2021}. Regardless of the X-ray production mechanism, most models describe the emission arising from a source into which preaccelerated electrons are injected, which is subsequently left to cool through radiative and/or adiabatic losses. It is therefore implicitly assumed that the acceleration of particles occurs instantaneously compared to all other relevant timescales.  

We present an acceleration-radiation model for the production of NIR and X-ray flares from an active region of the accretion flow. Our goal is to present an alternative interpretation for the NIR and X-ray flares of \sgras that is inspired by a model that was originally proposed for AGN flares \citep{KRM98}; see also \cite{MM2008}. We study the energization of electrons from their thermal pool to higher energies in a time-dependent way and account for their synchrotron and SSC emission. This is in contrast to other radiation models, in which the acceleration phase of the particles is usually not taken into account (because it is assumed to be instantaneous, e.g.).  Our study is motivated by multiwavelength observations of \sgras during bright flares, which show evidence for a later onset of the X-ray flare than in its NIR counterpart \citep{Ponti2017, GRAVITY2021}. An exciting possibility would be that we observing in real time particle energization in the vicinity of \sgras.  

This paper is structured as follows. In Sect.~\ref{sec:model} we introduce the flare model and provide analytical expressions for the electron distribution and magnetic field strength in the flaring region. We also describe the numerical code we used to calculate the radiative transfer. In Sect.~\ref{sec:results} we present numerical results for the time evolution of the spectral energy distribution (SED) and the NIR and X-ray light curves produced in our flare model for constant and time-variable particle injection into the acceleration process. In Sect.~\ref{sec:app} we apply our model to the 2014 NIR and X-ray flares of \sgras. We discuss our results in Sect.~\ref{sec:disc} and present our main conclusions in Sect.~\ref{sec:conc}. 

\section{Flare model}\label{sec:model}
\subsection{Model description}
We assumed that the nonthermal NIR and X-ray flares of \sgras are produced from an active region in the vicinity of the black hole (inner accretion flow or boundary between the funnel and the accretion disk), where particle acceleration takes place intermittently. The rising part of the flares is then attributed to the time when the acceleration is active and the decay when the accelerated particles are left to cool radiatively. 

The active region (or blob) is described as a spherical and homogeneous region of radius $R$ that contains relativistic electrons and a mean magnetic field of strength $B$. Particles with an initial Lorentz factor $\gamma_0 \gtrsim 1$ that enter the active region at a rate $Q_0$ accelerate to higher Lorentz factors on a timescale $t_{acc}$, while they can escape the acceleration region on a timescale $t_{esc}$. Particles also lose energy via synchrotron radiation. We limited our analytical calculations to cases in which inverse-Compton losses are subdominant, and thus, they cannot influence the electron distribution function. 

The temporal evolution of the particle distribution, $N_e(\gamma, t)$, is described by a partial differential equation (PDE) \citep{KRM98},
\begin{equation}
\frac{\partial N_e}{\partial t} + \frac{\partial}{\partial \gamma}\left( \left(\frac{\gamma} {t_{acc}} - b_s \gamma^2\right) N_e \right) + \frac{N_e}{t_{esc}} = Q_0 \delta(\gamma-\gamma_0),
\label{eq:kinetic}
\end{equation}
where $b_s = \sigma_T B^2 / (6 \pi m_e c)$, and $\delta(x)$ is the Dirac delta function. If $t_{acc}, t_{esc}$ are independent of the particle Lorentz factor\footnote{We discuss the effects of energy-dependent timescales in Appendix~\ref{app1:tacc-gamma}.}, and $Q_0$ is a constant, then Eq.~\ref{eq:kinetic} has a simple analytical solution, \citep{KRM98}
\begin{equation}
    N_e(\gamma, t) =  N_0 \gamma^{-s}\left( 1 - \frac{\gamma}{\gamma_{sat}}\right)^{s-2},\gamma_0 \le \gamma < \gamma_{\max}(t), 
    \label{eq:Ne}
\end{equation}
where $N_0 = Q_0 t_{acc} \gamma_0^{t_{acc}/t_{esc}}\left(1-\gamma_0/\gamma_{sat} \right)^{-t_{acc}/t_{tesc}}$, $s=1+t_{acc}/t_{esc}$, and 
\begin{equation} 
\gamma_{sat} = \left(b_s t_{acc}\right)^{-1} \propto B^{-2} t_{acc}^{-1},
\label{eq:gsat}
\end{equation}
is the saturation Lorentz factor (where the energy-loss rate balances the energy-gain rate due to acceleration). 
In Eq.~\ref{eq:Ne}, $\gamma_{\max}$ is the upper cutoff of the electron distribution, which evolves with time as
\begin{equation}
\gamma_{\max} (t) = \frac{\gamma_{sat}}{1+\left(\frac{\gamma_{sat}}{\gamma_0}-1\right) e^{-t/t_{acc}}}. 
\label{eq:gmax}
\end{equation}
We note that $\gamma_{\max}$ approaches the saturation Lorentz factor for $t \gg t_{acc}$ only asymptotically. 

Because $\gamma_{\max}$ increases monotonically with time during the acceleration phase, the synchrotron flux increases from longer to shorter wavelengths as the synchrotron cutoff frequency sweeps across the electromagnetic spectrum. If the acceleration acts upon particles indefinitely, the electron and photon distributions become stationary. In other words, a flare can be produced when the acceleration lasts for a finite time interval; this will become clearer in Sect.~\ref{sec:results}.

The optically thin synchrotron spectrum produced by electrons with $\gamma \ll \gamma_{sat}$ can be written as \citep{RL}
\begin{equation}
 \nu L_{syn}(\nu) = C(s) N_{e,tot} \gamma_{0}^{s-1} B^{(s+1)/2} \nu^{-s_{ph}+2}
 \label{eq:Lsyn},
\end{equation}
where $N_{e,tot} \approx \frac{Q_0 t_{acc}}{s-1}$ is the total number of electrons (at saturation)\footnote{The approximate relation is found by integrating Eq.~\ref{eq:Ne} over $\gamma$, ignoring the term in the parentheses, and using $\gamma_0 \ll \gamma_{sat}$.}, $C$ is a numerical constant that depends on $s$ as 
\begin{equation} 
C(s)= \frac{\sqrt{3}q^3 (s-1)}{m_e c^2 (s+1)} \left( \frac{2 \pi m_e c}{3 q} \right)^{-(s-1)/2} \Gamma\left( \frac{s}{4}+\frac{19}{12} \right) \Gamma\left( \frac{s}{4}-\frac{1}{12} \right),
\end{equation}
and  the synchrotron spectrum photon index is
\begin{equation} 
s_{ph}= \frac{s+1}{2} = 1+\frac{t_{acc}}{2t_{esc}}.
\label{eq:sph}
\end{equation}
If the latter is measured in the NIR band, for instance, then we can infer the ratio of the acceleration and escape timescales, $t_{acc}/t_{esc} = 2(s_{ph, \rm NIR}-1)$. 

In our model, the X-ray synchrotron radiation is produced by electrons with $\gamma \approx \gamma_{sat}$. We can then estimate the required magnetic field for electrons with $\gamma_{sat}$ to radiate synchrotron photons at a known photon energy $\epsilon_X$ as
\begin{eqnarray}
B & = &  t_{acc}^{-2/3} \left(\frac{m_e c^2}{\epsilon_X}\right)^{1/3}  \left(\frac{6 \pi m_e c}{\sigma_T B_{cr}^{1/2}} \right)^{2/3} \nonumber  \\
& \simeq & 1.2~{\rm G} \,  \left(\frac{2(s_{ph}-1)}{1.5}\right)^{-2/3} \left(\frac{R}{9.5 R_S} \right)^{-2/3}\left(\frac{\epsilon_X}{10~\rm keV}\right)^{-1/3}
\label{eq:B},
\end{eqnarray}
where $B_{cr} \simeq 4.4 \times 10^{13}$~G. The numerical value in the above equation was obtained for parameters that were also used in our numerical study in Sect.~\ref{sec:results}, namely $t_{acc}/t_{esc}=1.5$, $t_{esc} = R/c \simeq 400~{\rm s}$, for $R=9.5 R_S$, and $R_{S}=2GM/c^2$ is the Schwarzschild radius of a $4.3\times 10^6 M_\odot$ black hole \citep{GRAVITY2019}.  
 
Given the inferred magnetic field strength, we can estimate the cooling timescale of electrons radiating at NIR frequencies, $\nu_{\rm NIR}\sim 10^{14}$~Hz, 
\begin{equation}
t_{cool}(\nu_{\rm NIR}) \simeq 53.7~{\rm hr} \, \left(\frac{B}{1~\rm G} \right)^{-3/2} \left(\frac{\nu_{\rm NIR}}{10^{14}~\rm Hz}\right)^{-1/2}.
\label{eq:tcool}
\end{equation}
If the particles remain confined in the flaring region after the end of the acceleration phase, the long cooling timescale of NIR-emitting electrons would suggest an almost constant NIR flux over day-long timescales.  Meanwhile, the short cooling timescale of electrons emitting at 10~keV ($\sim 0.3$~hr for $B\sim 1$~G) would result in a fast decrease in the X-ray flux. In this case, the NIR and X-ray flares would have different decaying profiles. A decrease in the NIR flux on much shorter timescales than $t_{cool}(\nu_{\rm NIR})$ would require particle escape from the flaring region. In this case, the decaying profiles of NIR and X-ray flares would be similar, dictated by the photon-crossing timescale.

To model the decaying part of the flare, we assumed that the acceleration had ceased while particles were left to cool due to radiative losses and also escaped from the blob (similar to the radiation zone of the two-zone model of \cite{KRM98}). The kinetic equation for the decay phase of the flare is then written as
\begin{equation}
    \frac{\partial N^{\rm (II)}_{e}}{\partial t} + \frac{\partial}{\partial \gamma}(-b_s \gamma^2 N^{\rm (II)}_e) + \frac{N_e^{(II)}}{t^{(II)}_{esc}} = N_e\delta(t-t_{pk}),
    \label{eq:kinetic2}
\end{equation}
where we allowed for a different escape timescale of particles from the active region after the end of the acceleration episode by introducing the parameter $t_{esc}^{(II)} \ge t_{cr}$. We assumed that the acceleration ceased when $\gamma_{max}(t=t_{pk}) = \xi \gamma_{sat}$ with $\xi \lesssim 1$. Solving Eq.~\ref{eq:gmax} for $t_{pk}$, we find an estimate for the peak time of the flare (see also Sect.~\ref{sec:results}),
\begin{equation}
t_{pk} = t_{acc} \ln \left(\frac{\xi}{1-\xi} \left( \frac{\gamma_{sat}} {\gamma_0} -1 \right) \right).
\label{eq:tpk}
\end{equation}
For $t>t_{pk}$, Eq.~\ref{eq:kinetic2} can be solved using $N_e(\gamma, t=t_{pk})$ as an initial condition, where $N_e$ is given by Eq.~\ref{eq:Ne}, resulting in
\begin{equation}
    N^{\rm (II)}_e(\gamma, t> t_{pk}) = N_e\left(\frac{\gamma}{1-b_s\gamma (t-t_{pk})}, t \right) \frac{ e^{-(t-t_{pk})/t_{esc}} } {\left(1-b_s \gamma (t-t_{pk}) \right)^{2}}.
\end{equation}
In the postacceleration phase, adiabatic expansion of the blob and magnetic field decay could become relevant. Their impact on the flare decay profile has been investigated in detail in earlier studies \citep{Dodds-Eden_2010}. 

\subsection{Numerical approach}
We used the leptonic module of the numerical code {\sc athe$\nu$a} \citep{DMPR12}, which computes the temporal evolution of the lepton (electrons and positrons, $N_{e}(\gamma,t)$) and photon distributions $N_{ph}(\epsilon, t)$ (differential in energy)\footnote{Particle and photon energies are normalized to $m_e c^2$.} by solving a set of coupled PDEs (see below) using a backward-differentiation algorithm with a variable time step,

\begin{eqnarray}
\label{eq:pde1}
\frac{\partial N_e}{\partial t} + \frac{\partial}{\partial \gamma}\left(\frac{\gamma N_e}{t_{acc}}\right) +  \frac{N_e}{t_{esc}} \!\!\!\! &=& \!\!\!\! \mathcal{L}_e^{syn} + \mathcal{L}_e^{ics} + \mathcal{Q}_{e}^{\gamma \gamma} + Q_0 \delta(\gamma-\gamma_0) \\ 
\frac{\partial N_{ph}}{\partial t} + \frac{N_{ph}}{t_{esc}}\!\!\!\! &=& \!\!\!\! \mathcal{L}_{ph}^{\gamma \gamma} + \mathcal{L}_{ph}^{ssa} + \mathcal{Q}_{ph}^{syn} + \mathcal{Q}_{ph}^{ics} 
\label{eq:pde2},
\end{eqnarray}
where $t_{esc} = R/c$. The operators $\mathcal{L}_i^j$ and  $\mathcal{Q}_i^j$ denote the loss (sink terms) and production (source terms) rates of particle species $i$ due to the process $j$. The physical processes that are included in the code and couple photons with leptons are electron synchrotron emission, synchrotron self-absorption, electron inverse-Compton scattering, and $\gamma \gamma$ pair production. Finally, an acceleration term was added to the kinetic equation of electrons, according to Eq.~\ref{eq:kinetic}. This term is nonzero for a finite duration $T$, which is a free parameter. To model the decaying part of the flare, we set the acceleration term to zero for $t\ge T$.

The code takes as an input the electron compactness, $\ell_e$, which is the dimensionless measure of the injection power of electrons with $\gamma_0$,  and is defined as\footnote{This should not to be confused with the compactness, which is usually defined with respect to the bolometric power in nonthermal electrons.}
\begin{equation}
\ell_e = \frac{\sigma_T L^{\rm inj}_e}{4 \pi R m_e c^3} = \frac{\sigma_T Q_0 \gamma_0}{4 \pi R c},
\label{eq:comp}
\end{equation}
where $Q_0$ is the injection rate of electrons at $\gamma_0$. Given that $Q_0 \propto N_{e,tot}/t_{acc}$, the injection rate at $\gamma_0$ can be inferred from the observed flare luminosity (see Eq.~\ref{eq:Lsyn}) for a given $t_{acc}$. The numerical approach also offers the flexibility of studying time-variable injection rates, that is, $Q_0$ can depend explicitly on time (see Sect.~\ref{sec:q-var}).
 
After solving Eqs.~\ref{eq:pde1} and \ref{eq:pde2}, we can compute the escaping photon spectrum from the spherical blob of radius $R$ at time $t$ as $L(\epsilon, t) = 3 m_e c^2 \epsilon N_{ph}(\epsilon, t)/ t_{esc}$, where $L(\epsilon, t)$ is the escaping photon luminosity at time $t$ per photon energy $\epsilon$. Noting that $\epsilon = h\nu / (m_e c^2)$, we may write $L_{\nu}(t) = h L(\epsilon, t) / (m_e c^2)$, where $h$ is the Planck constant. The differential flux $F_\nu$ measured by an observer at a distance $D$ is then $F_{\nu}(t) = L_{\nu}(t)/(4 \pi D^2)$. The light curves are computed by integrating $F_\nu$ or $L_\nu$ in the desired energy/frequency range. These relations do not account for Doppler boosting or general relativistic effects, such as gravitational lensing and redshift. We discuss the effects of Doppler boosting on the light curves in a separate section (Appendix~\ref{app1:doppler}).

\begin{table}
\caption{Model parameters for an NIR and X-ray flare.}             
\label{tab:param}      
\centering      
\begin{threeparttable}
\begin{tabular}{l c c }
\hline 
Parameter & Symbol & Value \\ 
\hline 
Active region radius & $R$  & $1.2 \times 10^{13}~{\rm cm}$\tnote{*} \\ 
Light crossing time & $t_{cr}$ & 400 s \\
Particle escape timescale & $t_{esc}$  &  $t_{cr}$ \\ 
Acceleration timescale & $t_{acc}$ & $1.5 \, t_{esc}$ \\
Duration of acceleration episode & $T$ & $23 \, t_{acc}$ \\ 
Magnetic field strength & $B$ & 0.8~G \\ 
Initial electron Lorentz factor & $\gamma_0$ & $10^{2.7}$ \\ 
Electron compactness & $\ell_e$  & $10^{-3.5}$ \\
Electron injection rate at $\gamma_0$ & $Q_0$  & $3.2 \times 10^{42}$ s$^{-1}$ \\
\hline
\end{tabular}
\begin{tablenotes}
\item[*] $R\approx 9.5 R_S$ for a black hole mass of $4.3\times 10^6 M_\odot$ as for \sgras ~\citep{GRAVITY2019}.
\end{tablenotes}
\end{threeparttable} 
\end{table}

\section{Results}\label{sec:results}
We present multiwavelength photon spectra and light curves at NIR and X-ray energies computed numerically for an indicative set of parameter values listed in Table~\ref{tab:param}. We first considered the case of a constant injection rate of particles into the acceleration process, and then, we present the results for a stochastically variable injection rate.

\subsection{Constant injection rate}

\begin{figure*}
\centering
\includegraphics[width = 0.95\textwidth]{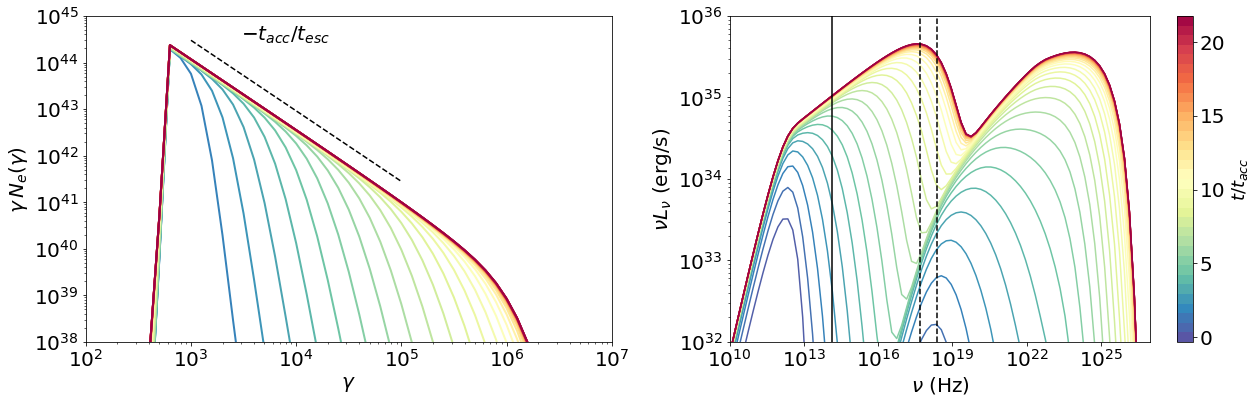}
\caption{Temporal evolution of the electron distribution $N_e(\gamma)$ multiplied by $\gamma$ (left panel) and of the emitted SSC spectra (right panel) during a particle acceleration episode for the parameters listed in Table~\ref{tab:param}. The colors indicate different times (in units of $t_{acc}$; see the color bar). The dashed line in the left panel shows the model-predicted slope for the electron distribution. The vertical solid and dashed lines in the right panel indicate the characteristic NIR (2.2~$\mu m$) and X-ray (2-10 keV) frequencies, respectively.}
\label{fig:sed-1}
\end{figure*}

\begin{figure}
\centering
\includegraphics[width = 0.4\textwidth]{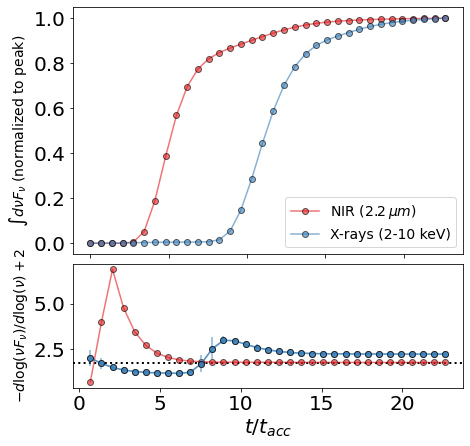}
\caption{Flux and photon index plotted as a function of time for the rising part of an NIR and X-ray flare. Top panel: NIR and X-ray light curves (normalized to their peak values) of the SSC model presented in Fig.~\ref{fig:sed-1}. Bottom panel: Temporal evolution of the photon index at NIR and X-ray frequencies. For the latter case, the average value of the photon index in the 2-10 keV band is plotted.  The error bar shows the standard deviation of the values in the 2-10 keV range. The horizontal dotted line marks the theoretical value of the photon index.  In both panels, the markers are used to show the cadence of the numerical calculation ($1\, t_{cr}$).} 
\label{fig:lc-1}
\end{figure}

Fig.~\ref{fig:sed-1} (left panel) shows the temporal evolution of the electron distribution, $N_e(\gamma)$, during an acceleration episode lasting $20 \, t_{acc}$. A power law is formed as the high-energy cutoff of the distribution increases until it reaches its saturation value ($\gamma_{sat} \approx 10^{6.2}$). We note that the particle distribution has no cooling break, as is expected when particles sustain only radiative energy losses upon their injection into the emitting region. The emitted synchrotron and SSC  spectra are displayed on the right. Due to the progressive build-up of the power-law electron distribution, the model predicts a lag between the onset of the flare at NIR and X-ray frequencies. This is illustrated more clearly in the upper panel of Fig.~\ref{fig:lc-1}, where the light curves at 2.2 $\mu m$ (NIR) and 2-10 keV (X-rays) are plotted. The light curves are normalized to their peak values to facilitate comparison. In this example, the X-ray flux reaches 80 per cent of its peak value at about $5 \, t_{acc}$ after the NIR flux. The model also predicts a strong spectral evolution during the rise of the light curves, which is caused by the passage of the synchrotron cutoff frequency through the respective energy band (see the lower panel in Fig.~\ref{fig:lc-1}).

\begin{figure}
\centering
\includegraphics[width=0.45\textwidth]{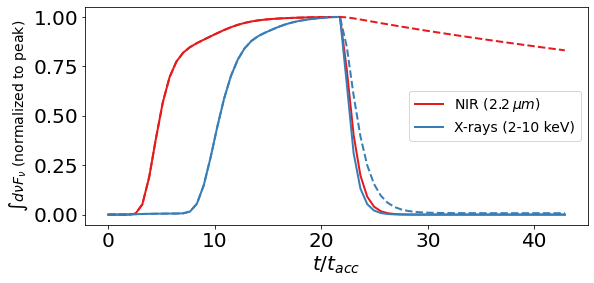} 
\includegraphics[width=0.45\textwidth]{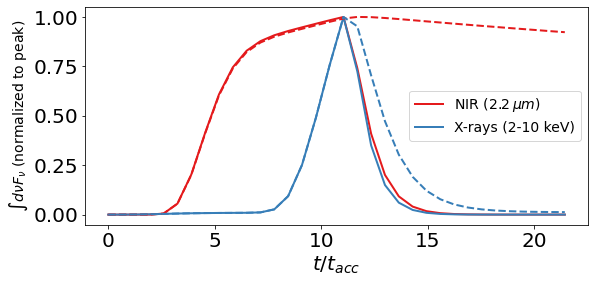}
\caption{Light curves of an NIR and X-ray flare for two different choices of the particle acceleration duration. Top panel: NIR and X-ray light curves for the same flare as in Figs.~\ref{fig:sed-1} and \ref{fig:lc-1}, but also accounting for the decay phase. We show the results for two choices of the particle escape timescale during the decay: $t_{esc}^{(II)}=R/c$ (solid lines) and $t_{esc}^{(II)}\rightarrow \infty$ (dashed lines). 
Bottom panel: Same as in the top panel, but for an acceleration episode lasting $10 \, t_{acc}$.
}
\label{fig:lc-decay}
\end{figure}

After the end of the acceleration episode, particles are left to cool and escape from the active region on the respective timescales (see also Sect.~\ref{sec:model}). Depending on the particle escape timescale, different decaying flare profiles can be obtained. This is illustrated in the top panel of Fig.~\ref{fig:lc-decay}, where we show results for two extreme cases: Particles escape on $R/c$, that is, on the shortest possible timescale from the flaring region  after the end of the acceleration phase (solid lines), or they are not allowed to escape at all (dashed lines). In the latter case, the NIR flux decays on the cooling timescale of the radiating electrons, which is much longer than that of X-ray emitting electrons (see also Eq.~\ref{eq:tcool}). Moreover, the X-ray profile is asymmetric in both cases, and the rise time is longer than the decay time of the flare because a plateau forms in the X-ray light curve. However, when the acceleration episode does not last long enough to lead to a saturation of the high-energy cutoff of the electron distribution, the X-ray flare will appear symmetric, as shown in the bottom panel of Fig.~\ref{fig:lc-decay}. Finally, the flare profiles at longer wavelengths are generally asymmetric, unless the particle escape timescale $t_{esc}^{(II)}$ is set to be comparable to the rise time. 

In Appendix~\ref{app1} we discuss how other model parameters, such as the initial Lorentz factor of electrons before their injection into the acceleration phase and the size of the active region, may affect the multiwavelength spectra of the flare. Moreover, we discuss the case of the energy-dependent acceleration/escape timescales and the impact of Doppler boosting on the shape of the light curves, which may arise due to the relative motion of the active region and the observer by considering an indicative trajectory. 

\subsection{Variable injection rate}\label{sec:q-var}
\begin{figure*}
\centering
\includegraphics[width = 0.9\textwidth]{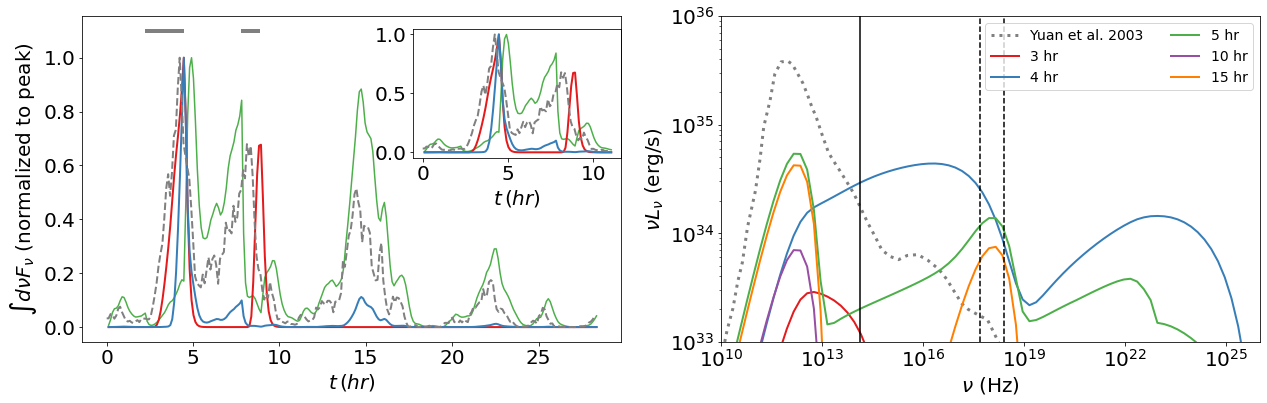}
\caption{Light curves and spectra of a nonthermal flare computed for a variable particle injection rate. Left panel: Light curves at different energies (500 GHz: green; NIR: red; X-rays: blue) computed for a variable injection rate $Q_0$ that mimics red noise (dashed gray line). All other parameters are the same as in Table~\ref{tab:param}. Two episodes of acceleration are assumed to take place and are indicated by the horizontal bars. The time series in the left panel are normalized to their peak values, and a zoom in the first 10 hr is shown in the inset plot.  The light curves do not account for the background flux arising from the accretion flow. Right panel: SED snapshots of our model. For comparison, the quiescent emission model of \citet{Yuan2003} is also shown (dotted curve). An animation of the full temporal evolution of the broadband spectra can be found at this \href{https://github.com/mariapetro/SgrA-flares/blob/main/sed-q_var4-new.gif}{link}.}
\label{fig:lc-sed-var}
\end{figure*}

In the previous section, we assumed that the injection rate of particles with $\gamma_0$ was constant during the flare. To demonstrate the effects of variable injection of low-energy electrons into the flaring region, we constructed a red-noise time series for $Q_0$\footnote{The time resolution of the series is $R/c$.} using the Python package {\tt colorednoise}. This is shown by the dashed grey line in the left panel of Fig.~\ref{fig:lc-sed-var}. We also introduced two acceleration episodes that lasted two hours and one hour each, as indicated by the horizontal lines in the same figure. All other parameters were the same as those listed in Table~\ref{tab:param}. In the first acceleration episode, lasting about $20 \, R/c$, the high-energy cutoff of the electron distribution approached $\gamma_{sat}$. In contrast, $\gamma_{\max}$ in the second episode was limited by the shorter duration to lower values. 

In Fig.~\ref{fig:lc-sed-var} we present (normalized) light curves (left panel) and SED snapshots of the emission produced by nonthermal electrons in the flaring region. In the absence of acceleration, electrons with $\gamma_0 \sim 500$ produce $\sim 500$~GHz photons (green line) in a region with a magnetic field of strength of $\sim 1$~G, as shown in the first SED snapshot in the right panel (at 3~hr). Hence,  before the start of the first acceleration episode, no NIR (2.2 $\mu m$) emission is expected from the active region. Soon after the onset of the first acceleration episode, the NIR (red line) and X-ray (blue line) fluxes rise sharply, and the X-rays lag about $\sim 5 t_{acc} \sim 0.8$~hr with respect to the NIR frequencies (see also Fig.~\ref{fig:lc-1}). The amplitude of the flux variations in the X-rays is expected to be larger than in NIR frequencies. During the first acceleration episode, the flux at 500 GHz is suppressed and does not follow the injection rate profile because particles injected at $\gamma_0$ are shifted to higher energies (see the inset plot in the left panel). In other words, the injected energy that would otherwise be radiated at 500 GHz is now channeled to higher frequencies. Interestingly, similar indications have been reported by \cite{2022ApJ...930L..19W}, who found spectral changes at 1.3~mm close to the peak time of an X-ray flare in \sgras (see Fig.~6 in their paper). At the end of the acceleration phase, the NIR and X-ray fluxes decrease on a light-crossing timescale (because the electrons in this example escape from the region on the same timescale), while the 500 GHz flux increases (see also SEDs at 4 hr and 5 hr in the right panel).

\begin{figure*}
    \centering
    \includegraphics[width=0.9 \textwidth]{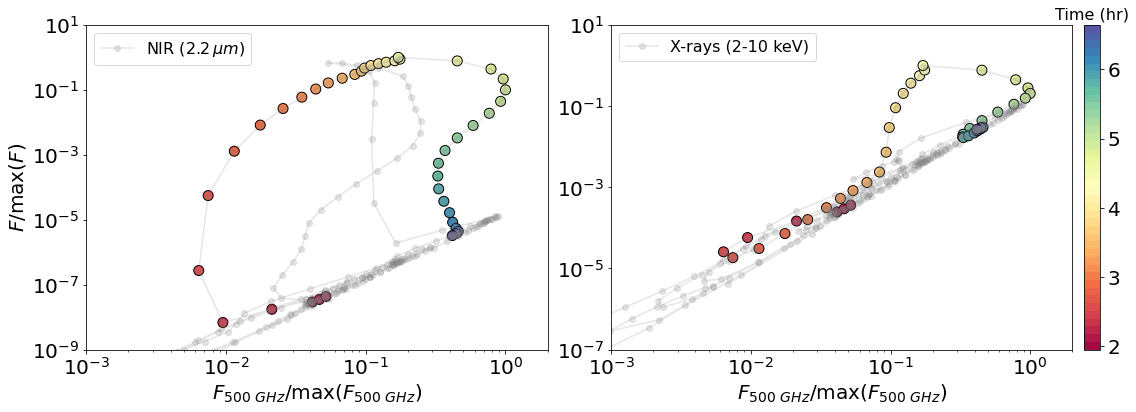}
    \caption{Flux-flux diagram for the case presented in Fig.~\ref{fig:lc-sed-var} (gray symbols). The colored symbols indicate the evolution during the first nonthermal flare shown in Fig.~\ref{fig:lc-sed-var} (see also the color bar). All fluxes refer to the flaring region.}
    \label{fig:flux-flux}
\end{figure*}

The second acceleration episode is identified by a rapid decrease in the 500 GHz luminosity and the emergence of an NIR flare, which lacks an X-ray counterpart, however: Because this episode is shorter, particles injected with $\gamma_0$ do not have enough time to reach a large enough Lorentz factor, and they emit X-rays via synchrotron radiation. Still, electrons at $\gamma_0$ produce SSC radiation at a characteristic frequency $\nu_{ssc} \approx 1.7\times 10^{17}~{\rm Hz} \, (\gamma_0/500)^2 (\nu_{syn}/500~\rm GHz)$, which emerges in the X-rays for the chosen parameters. Therefore, enhancements in the X-ray luminosity that are not accompanied by NIR flares are also expected due to the stochastic changes in the electron injection rate (see, e.g., the SEDs in the right panel at 15 hr). In general, the X-ray luminosity in this case is lower than the luminosity expected from a synchrotron flare caused by particle acceleration (compare the X-ray luminosities at $\sim 5$~hr and 15 hr). 

In Fig.~\ref{fig:flux-flux} we plot the X-ray and NIR fluxes (normalized to their peak values) against the 500 GHz flux from the flaring region that tracks the variations in the injection rate. The colored markers indicate the flux evolution from the period of the first acceleration episode, lasting about 4~hr. NIR and X-ray flares produced by an episode of particle acceleration that is followed by a particle cooling/escape phase form clockwise loops in this diagram. The later onset of the X-ray flare results in a smaller loop than for the NIR flare. Loops in flux-flux diagrams have long been sought for in AGN X-ray and TeV $\gamma$-ray flares \citep[e.g.][]{MM2008, MM2011} because they can provide a glimpse into the underlying particle acceleration process.

\begin{figure*}
\centering
\includegraphics[width=0.47\textwidth]{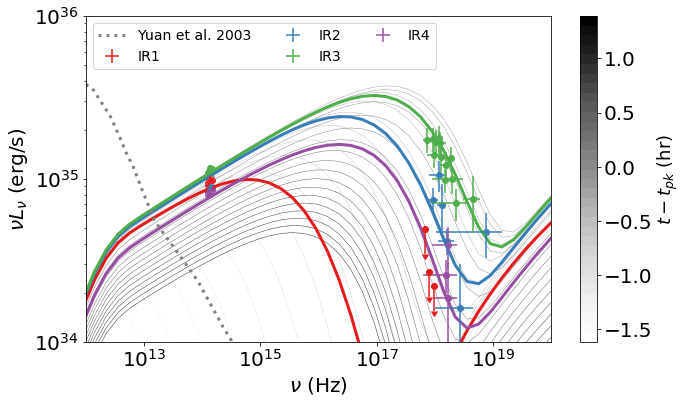} 
\includegraphics[width=0.47\textwidth]{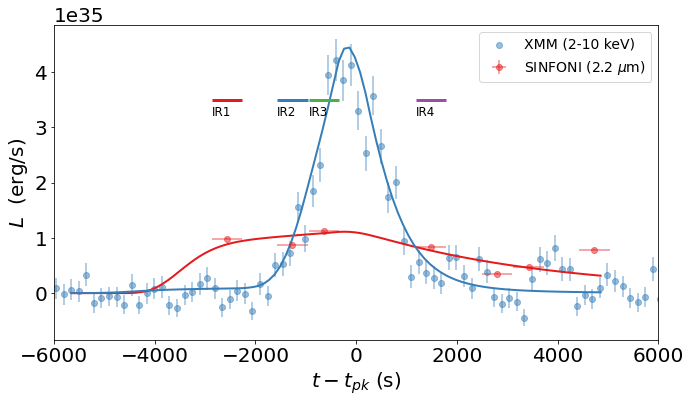}
\caption{Application of the model to the 2014 nonthermal flare from \sgras. Left panel: SED evolution during the 2014 bright flare of \sgras (symbols). Snapshots of model spectra (every $t_{cr}$ is plotted with thin gray lines). The thick lines represent the average model spectra during the four observation time windows. Right panel: NIR and X-ray model light curves (solid lines) plotted against the data of the 2014 flare (symbols). After the peak time of the flare, the particles escape from the active region on a timescale $12.5 \, R/c$. The data are adopted from \citet{Ponti2017}.}
\label{fig:flare_2014}
\end{figure*}

\begin{figure}
\centering
\includegraphics[width=0.45\textwidth]{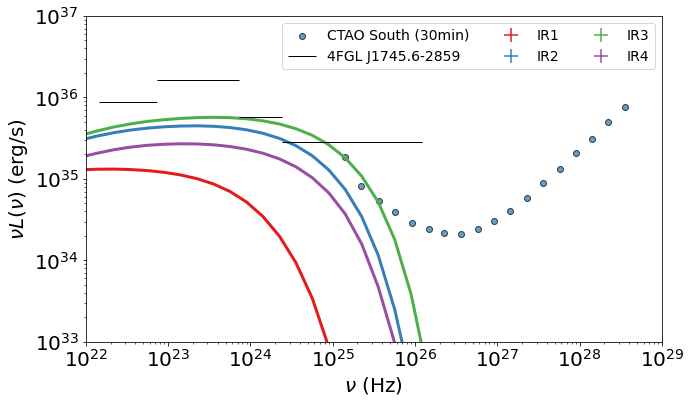}
\caption{Zoom in the $\gamma$-ray range of the model spectra computed for epochs IR1-IR4. The differential sensitivity curve of CTAO South for an observing time of 30~min is overplotted. The black lines indicate the 10.5 yr integrated Fermi-LAT spectrum for the $\gamma$-ray source 4FGL J1745.6-2859, which is associated with the Galactic center~\citep{Cafardo_2021}.} 
\label{fig:flare2014_cta}
\end{figure}

\section{Application to the 2014 NIR and X-ray flare}\label{sec:app}
We considered the first simultaneous observations of X-ray and NIR emission of a very bright flare of \sgras, which occurred between 2014 August 30 and 31  \citep{Ponti2017}. We applied our model to four time windows identified by \citet{Ponti2017} in which simultaneous NIR and X-ray spectra could be constructed. These epochs, each one lasting 600~s, correspond to the pre-rise (IR1), rise (IR2), peak (IR3), and decay (IR4) of the flare. We reprocessed the XMM-Newton data using the same procedure as \cite{Ponti2017}, but with the 2023 calibration files. The data are fully consistent with those reported in \citet{Ponti2017}.

Having as guideline the analytical expressions in Sect.~\ref{sec:model}, we find physically motivated parameter values that can reproduce the general spectral and temporal properties of the observed flare. Our results, which were obtained for similar parameter values as in the default set used in the previous section (see Tables~\ref{tab:param} and \ref{tab:flare_2014}), are presented in Fig.~\ref{fig:flare_2014}.  The model naturally explains the X-ray nondetection during IR1 because the highest-energy particles of the distribution have not yet reached high enough energies to emit in the X-rays. To reproduce the fast and slow decays of the X-ray and NIR flares, respectively, a longer escape timescale ($12.5 R/c$) from the active region is required after the acceleration ceases to operate. While the fast decay of the X-ray flare is dictated by the synchrotron cooling timescale, the NIR decay timescale reflects the electron escape rate. 

In addition to the SED of the nonthermal flare, we show in Fig.~\ref{fig:flare_2014} the spectrum of the accretion flow based on the model of \cite{Yuan2003}. If the flare is produced in a compact region of the accretion flow (blob), it could be embedded in a photon bath produced by the hot electrons in the flow. These low-energy photons could provide external targets for inverse-Compton scattering by the electrons in the blob. The energy density of the submm quiescent photon field is $u_{smm} = L_{smm}/(4\pi R_{smm}^2 c) \simeq 6\times 10^{-3} \, L_{smm, 35.6} (10 R_S/R_{smm})^2$~erg cm$^{-3}$, where we used the peak luminosity of the submm bump as a proxy for the bolometric luminosity. The magnetic energy density of the blob producing the flare is $u_B \simeq 4\times 10^{-2} (B/ 1{\rm G})^2$~erg cm$^{-3}$. Since $u_B > u_{smm}$, inverse-Compton scattering in the submm quiescent photon field contributes significantly to the high-energy spectrum of the flare (and to the cooling of nonthermal electrons).

\begin{table}
\caption{Model parameter values for the 2014 bright flare of Sgr A*.}             
\label{tab:flare_2014}      
\centering      
\begin{threeparttable}

\begin{tabular}{l c c }
\hline 
Parameter & Symbol & Value \\ 
\hline 
Active region radius & $R$  & $9\times10^{12}$~cm \\ 
Light crossing time & $t_{cr}$ & 304 s \\
Particle escape timescale & $t_{esc}$  &  $t_{cr}$ \\ 
Particle escape timescale (II)\tnote{\textdagger} & $t^{(II)}_{esc}$  &  $12.5 \, t_{cr}$ \\
Acceleration timescale & $t_{acc}$ & $1.5 \, t_{esc}$ \\
Duration of acceleration episode & $T$ & $16 \, t_{cr}$ \\ 
Magnetic field strength & $B$ & 0.8~G \\ 
Initial electron Lorentz factor & $\gamma_0$ & $10^{2.7}$ \\ 
Electron compactness & $\ell_e$  & $10^{-3.2}$ \\
Electron injection rate at $\gamma_0$ & $Q_0$  & $6.4 \times 10^{42}$ s$^{-1}$ \\
\hline
\end{tabular}
\begin{tablenotes}
\item[\textdagger]  After the acceleration episode.
\end{tablenotes}
\end{threeparttable} 
\end{table}

So far, no NIR and X-ray flare from \sgras has been observed in very high energies (VHE; $>100$~GeV). In Fig.~\ref{fig:flare2014_cta} we show a zoom in the $\gamma$-ray spectra of our 2014 flare model. The cutoff of the SSC spectrum at the peak time of the X-ray flare (IR3) scratches the 30 min sensitivity curve\footnote{The adopted exposure is comparable to the full width at half maximum of the X-ray flare (see Fig.~\ref{fig:flare_2014}).} of the Cherenkov Telescope Array Observatory (CTAO) South \citep{CTA_2021}.  CTAO South could place meaningful upper limits or even secure a detection of the VHE counterpart of bright X-ray flares from \sgras in the future. We also overplot the 60~MeV to 500~GeV spectrum of the point source 4FGL J1745.6-2859, which was constructed using 10.5~years of Fermi-LAT data, and is thought to be associated with \sgras \citep{Cafardo_2021}. However, a GeV counterpart of the flare would remain undetectable by Fermi-LAT due to the small fluence (short duration of the flares). 

\section{Summary and discussion}\label{sec:disc}
We have presented a model for the production of NIR and X-ray flares motivated by multiwavelength observations of \sgras. The model invokes a phase of particle energization that is followed by a period of particle cooling and escape from the active region. We first provided an analytical description of the model that highlighted the role of key physical parameters and their relation to flare observables. We then presented numerical results for the flare SED and light curves and applied our model to the 2014NIR and X-ray flare of \sgras.

There are several differences between our model and the standard synchrotron cooling model for \sgras flares regarding the physical conditions in the flaring region that are worth mentioning. For NIR and X-ray flares with similar characteristics as those of the 2014 flare, our model predicts much weaker magnetic fields ($\sim 1$~G) and more extended flaring regions ($R \sim (5 - 10) \, R_S)$ than the synchrotron cooling model~\citep{Ponti2017, GRAVITY2021}. An attempt to consider smaller radii to explain the same NIR and X-ray luminosity would result in a flare with strong Compton dominance, as demonstrated in Appendix~\ref{app1:radius}, which would contradict the soft X-ray spectrum. Moreover, the spectral break between the NIR and X-ray energy ranges in our model is not related to the cooling of particles, but to the onset of the exponential cutoff of the synchrotron spectrum. The evolution of this spectral cutoff, $\nu_{\max} \propto B \gamma_{\max}^2$, is also very distinct: It increases exponentially during the rise of the flare, namely  $\nu_{\max} \propto e^{2t/t_{acc}}$ (see Eq.~\ref{eq:gmax}), and decreases during the decay as a power law with time, $\nu_{\max}\propto (1+ b_s \gamma_{sat}(t-t_{pk}))^{-2}$. Last but not least, our model predicts a lag between the onset of the X-ray and NIR flares (see, e.g., Fig.~\ref{fig:lc-decay}), which can be directly linked to the particle acceleration timescale. In contrast, the flux in both energy bands evolves simultaneously in the synchrotron cooling-break model. 

In our framework, NIR flares without an X-ray counterpart are
naturally expected when the duration of an acceleration episode is shorter than the time needed to accelerate particles to sufficiently high energies (see Figs.~\ref{fig:lc-sed-var}). Using analytical expressions from Sect.~\ref{sec:model}, we can estimate the number ratio of NIR flares with and without X-ray emission predicted by our model. Let $\gamma_X=\gamma_{sat}$ be the Lorentz factor of electrons emitting X-ray photons of energy $\epsilon_X$. Using Eqs.~\ref{eq:gmax} and \ref{eq:B}, we find that $\gamma_X = t_{acc}^{1/3}(m_e c^2/\epsilon_X)^{-2/3}(6\pi m_e c /\sigma_T)^{-1/3}B_{cr}^{2/3}$. Here, $t_{acc} = 2 t_{esc}(s_{ph, \rm NIR}-1)$, where $s_{ph, \rm NIR}$ is the NIR to X-ray photon index.  Similarly, the Lorentz factor of electrons emitting at NIR can be written as  $\gamma_{NIR} = t_{acc}^{1/3}(m_e c^2/\epsilon_{NIR})^{-1/2} (m_e c^2/\epsilon_X)^{-1/6}(6\pi m_e c /\sigma_T)^{-1/3}B_{cr}^{2/3}$. The time needed for particles with an initial Lorentz factor $\gamma_0$ to reach $\gamma_X, \gamma_{NIR}$ can be obtained from Eq.~\ref{eq:tpk}, namely $t_{X,NIR}=t_{acc} \ln\left[\xi/(1-\xi) (\gamma_{X, NIR}/\gamma_0-1)\right]$. The ratio of the two timescales is $t_{X}/t_{NIR}\sim 1.6$ for the parameters used in Fig.~\ref{fig:lc-1}, but it does not depend strongly on the model parameters because of the logarithmic term. Assuming that the duration of the particle acceleration phase, $T$, follows a power-law distribution, ${\rm d}N/{\rm d}T \propto T^{-a}, a>1$, we can estimate the number of NIR without an X-ray counterpart as $\int_{t_{NIR}}^{t_X} {\rm d}T \, {\rm d}N/{\rm d}T \propto t_{NIR}^{-a+1}-t_{X}^{-a+1}$. The number of flares expected in both NIR and X-rays is $\int_{t_X}^\infty {\rm d}T \, {\rm d}N/{\rm d}T \propto t_{X}^{-a+1}$.  Therefore, the number ratio of NIR and X-ray flares to NIR flares is $N_{NIR/X}/N_{NIR} = (t_{X}/t_{NIR})^{-a+1}/[ 1- (t_{X}/t_{NIR})^{-a+1}] $, which ranges between 0.2 and 0.8 for $a=5$ and $a=3$, respectively. These estimates are optimistic because they do not account for instrumental effects and observational biases. 

So far, we have remained agnostic to the nature of the acceleration mechanism. GRMHD simulations of accretion onto black holes have shown the formation of current sheets (in the disk and in the boundary between the disk and the funnel) that are potential sites of energy dissipation and particle acceleration through magnetic reconnection \citep[e.g.][]{Ball_2018_mhd, nathanail2020, Nathanail_2022, Ripperda_2022}. The properties of the accelerated particles (e.g., power-law slope and maximum energy) depend on local plasma properties, such as the magnetization $\sigma$ (defined as twice the ratio of the magnetic energy density to the plasma enthalpy density), the plasma-$\beta$ parameter (defined as the ratio of thermal and magnetic pressures), and the plasma composition \citep[e.g.][]{2014ApJ...783L..21S, Ball_2018_pic, 2018MNRAS.473.4840W,Petropoulou_2019}. During reconnection, particles may gain energy by the reconnection electric field $E_{rec} \sim \beta_{rec} B$ on a timescale $t_{acc} \approx m_e \gamma c^2/ (e \beta_{rec} B c)$, where $\beta_{rec}\sim 0.1$ is the reconnection rate and $B$ is the strength of the reconnecting magnetic field \citep[for 3D simulations, see also][]{Zhang2021, 2023ApJ...956L..36Z}. The saturation Lorentz factor in this scenario can be estimated by balancing the synchrotron cooling and acceleration timescales, namely $\gamma_{sat} = (6 \pi e \beta_{rec} /(\sigma_T B))^{1/2} \simeq 3.6\times 10^7~(\beta_{rec}/0.1)^{1/2} (B/1~\rm G)^{-1/2}$. Electrons with $\gamma_{sat}$ therefore radiate synchrotron photons of $\sim 15.8$~MeV energy, regardless of the magnetic field strength \citep[synchrotron burnoff limit, ][]{deJager1996}. Because there is evidence of a spectral cutoff at tens of keV in X-rays flares from \sgras (see, e.g., Fig.~\ref{fig:flare_2014}), we can conclude that the maximum electron energy cannot be limited by radiation. It could instead be limited by the size $\ell$ of the accelerator \citep[Hillas limit,][]{1984ARA&A..22..425H}, which can be estimated as $\ell = m_e c^2 \gamma_{X}/e E_{rec}  = m_e c^2 (\epsilon_X m_e c/\hbar)^{1/2} \beta_{rec}^{-1} (e B)^{-3/2} \simeq 0.02\, R_S \, (\epsilon_X/20~{\rm keV})^{1/2} (\beta_{rec}/0.1)^{-1} (B/1~{\rm G})^{-3/2}$. To compensate for the small emitting volume while accounting for the observed luminosity of flares, a large number of nonthermal electrons would be needed. The high-electron compactness of the emitting region would lead to Compton-dominated SEDs, which is in contrast to observations (see, e.g., Fig.~\ref{fig:sed-2}). Moreover, the acceleration timescale is very short compared to the typical duration of X-ray flares for typical magnetic field strengths in the inner accretion flow, for example, $t_{acc}\approx 10^{-6} \, (\gamma/100) (B/1 {\rm G})^{-1}~(R_S/c)$. As a result, the acceleration would be instantaneous, and neither the rise time of the X-ray flare nor a lag between NIR and X-ray flares could be explained by $t_{acc}$ in this case. Therefore, if direct acceleration by the reconnection electric field in macroscopic current sheets energizes particles during flares in \sgras, then models considering the injection of preaccelerated particles into a radiation zone in which particles cool and/or escape would be appropriate for the flare description \citep[e.g.][]{Dodds-Eden_2010, Ball_2021, 2024MNRAS.531.3136L, 2024arXiv240714312D}. 

Alternatively, particles can accelerate in environments of magnetized turbulence by the reconnecting electric field at transient current sheets and/or stochastically through multiple scatterings by turbulent fluctuations. Recently, particle-in-cell (PIC) simulations of magnetized turbulence have demonstrated from first principles the formation of power-law particle distributions \citep[][]{Zhdankin_2017, Zhdankin_2018, Comisso_2018}. By analyzing particle trajectories in PIC simulations, \cite{Comisso_2019} showed that low-energy particles typically experience a small energy gain at the current sheets before they are injected into a diffusive acceleration process. Interestingly, the authors found that the relevant acceleration timescale is independent of energy (nongyroresonant scatterings; see also \cite{Wong_2020}) and can be expressed in terms of the turbulence driving-scale $\ell_t$ and of the plasma magnetization (defined with respect to the turbulent component of the magnetic field) $\sigma_{t}$,  as $t_{acc} \approx (10/\sigma_{tur}) \ell_t/c$ \citep[see also][]{Fiorillo_2024arXiv}. Writing $\ell_t$ as a fraction $\eta$ of the source size $R$, we find that $t_{acc} \approx t_{cr} (\eta/0.1) (\sigma_t/1)^{-1}$, which is comparable to the value used to model the 2014 NIR and X-ray flare (see Table~\ref{tab:flare_2014}). Furthermore, the synchrotron-limited maximum particle energy is also similar to the energy found in our generic model (see Sect.~\ref{sec:model}). Despite these promising similarities, a more detailed investigation of the nonresonant acceleration-radiation model would require us to solve the time-dependent kinetic equation of Eq.~\ref{eq:kinetic} after replacing the second term with a diffusive one, $t^{-1}_{acc}\left(\gamma^2\frac{\partial^2 N_e}{\partial^2 \gamma} - 2 N_e\right) $. The new PDE admits stationary solutions of the form $N_e \propto \gamma^{-1}$ (power law) for $\gamma \ll \gamma_{sat}$, and $N_e \propto \gamma^2$ for $\gamma \lesssim \gamma_{sat}$ (pile up) \citep[for more details, see][]{Fiorillo_2024arXiv}. Compared to the stationary solution of our generic model, the main difference is the formation of a pile-up, which may result in interesting observable features.

\section{Conclusions}\label{sec:conc}
We have introduced a generic acceleration-radiation model for nonthermal flares in \sgras. According to the model, particles are energized in an active region with a size of a few Schwarzschild radii on an energy-independent acceleration timescale while experiencing mainly synchrotron losses. Particles physically escape this region on a timescale comparable to their acceleration timescale. The emitted synchrotron spectra are power laws, with a photon index determined by the ratio of the acceleration and escape timescales, followed by an exponential cutoff. This occurs at the synchrotron photon energy emitted by particles with the maximum Lorentz factor (where the energy loss and energy gain rates become equal). The model predicts an increases in NIR flux before the onset of the X-ray flare. This time difference as well as the rise time of the X-ray flare are multiples of the particle acceleration timescale. Our generic model for NIR and X-ray flares favors an interpretation of diffusive (nonresonant) particle acceleration in magnetized turbulence. This requires a more detailed investigation.

\begin{acknowledgements}
 M.P. acknowledges support from the Hellenic Foundation for Research and Innovation (H.F.R.I.) under the ``2nd call for H.F.R.I. Research Projects to support Faculty members and Researchers" through the project UNTRAPHOB (Project ID 3013).  This work is based on observations obtained with XMM-Newton, an ESA science mission with instruments and contributions directly funded by ESA Member States and NASA. G.P. acknowledges financial support from the European Research Council (ERC) under the European Union’s Horizon 2020 research and innovation program HotMilk (grant agreement No. 865637), support from Bando per il Finanziamento della Ricerca Fondamentale 2022 dell’Istituto Nazionale di Astrofisica (INAF): GO Large program and from the Framework per l’Attrazione e il Rafforzamento delle Eccellenze (FARE) per la ricerca in Italia (R20L5S39T9).   
\end{acknowledgements}
%
  \bibliographystyle{aa} 
  \bibliography{flares.bib} 
%
\appendix 
\section{Effects of model parameters}\label{app1}
In this section we demonstrate the effect of model parameters, like the initial Lorentz factor of electrons before their injection into the acceleration phase, and the size of the active region, on the multi-wavelength spectra of the flare. We also discuss a scenario with energy-dependent particle timescales, and investigate how the relative motion of the active region and the observer can affect the shape of the light curves. 

\subsection{Size of flaring region}\label{app1:radius}
In the acceleration-radiation model for flares the ratio $t_{esc}/t_{acc}$, which determines the asymptotic power-law slope of the particle distribution, can be inferred by the observed photon index at NIR frequencies. Given the latter, a smaller flaring region would imply shorter $t_{acc}$ as long as $t_{esc} = R/c$. Because the acceleration would become faster, a stronger magnetic field, $B \propto t_{acc}^{-2/3} \propto R^{-2/3}$, would be needed to stop the acceleration at a maximum electron energy that will be sufficient to produce X-ray synchrotron photons -- see Eq.~(\ref{eq:B}). Moreover, if the 
electron injection compactness is fixed, then the 
injection rate of electrons at $\gamma_0$ is $Q_0 \propto R$ according to Eq.~(\ref{eq:comp}). Therefore, the synchrotron luminosity of a flare (at a fixed frequency $\nu$) produced from a smaller region would decrease as $\nu L_{syn}(\nu) \propto R^{4/3 - t_{acc}/(3 t_{esc})}$ -- see Eq.~(\ref{eq:Lsyn}).  

\begin{figure}
\centering
\includegraphics[width = 0.45\textwidth]{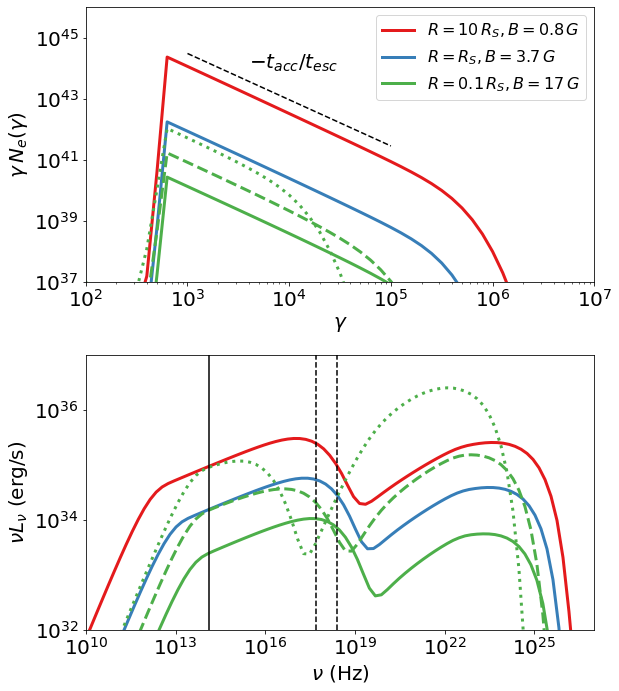}
\caption{Effects of the flaring region size on particle and photon spectra of a nonthermal flare. Top panel: Electron distribution $N_e(\gamma)$ multiplied by $\gamma$ at $t=10 \, t_{acc}$ for different radii and magnetic field strengths of the flaring region (see inset legend) and the same electron compactness $\ell_e$ (see Table~\ref{tab:param}). For $R=0.1 \, R_S$ we also show results for $5 \ell_e$ (dashed line) and $25 \ell_e$. Bottom panel:  SSC spectra produced by the electron distributions shown on the left panel.}
\label{fig:sed-2}
\end{figure}

These effects are illustrated in Fig.~\ref{fig:sed-2} where we plot (solid lines) the electron distributions (top panel) and photon spectra (bottom panel) at $t = 10 \, t_{acc}$ for $R={10, 1, 0.1} \, R_S$. In all cases, the magnetic field $B$ has been adjusted accordingly as to produce a synchrotron cutoff energy falling in the 2-10 keV range (see inset legend). Meanwhile, the dynamic range of the power-law synchrotron spectrum decreases as $B$ gets stronger and $\gamma_0$ remains the same. Moreover, the synchrotron luminosity decreases by a factor of $\sim 0.15$ as the radius decreases by a factor of 10, in agreement with Eq.~(\ref{eq:Lsyn}) -- see also scaling relation at the end of the previous paragraph. To achieve the same synchrotron luminosity in all three cases, the injection rate $Q_0$ would have to increase accordingly. This is exemplified for the case with $R = 0.1 R_S$ (see dashed and dotted green lines). As the particle density increases, SSC cooling becomes progressively more important than synchrotron, thus pushing the synchrotron cutoff energy below the X-ray band, and increasing the luminosity of the SSC component (i.e. Compton-dominated flare). 

Using the analytical expressions in Sect.~\ref{sec:model} we can also predict the radiative output of flaring regions with the same magnetic field but different radii. For example, if $B$ is constant and $Q_0 \propto R$, then the synchrotron luminosity of a flare (at a fixed frequency $\nu$) would scale as $\nu L_{syn}(\nu) \propto R^2$, while the cutoff synchrotron frequency would scale as $\nu_{sat} \propto B \gamma_{sat}^2 \propto t_{acc}^{-2} \propto R^{-2}$. Hence, smaller regions would produce less luminous NIR and X-ray flares, with synchrotron spectra extending beyond the X-ray band. Alternatively, if one considers that the power injected into the electrons with $\gamma_0$ is the same regardless of the flaring region size, i.e. $Q_0$ does not depend on $R$, the luminosity of the flare would scale as  $\nu L_{syn}(\nu) \propto R$. 

Summarizing, most model parameters ($t_{acc}/t_{esc}, B, R$) can be constrained if the photon index, the luminosity and spectrum of the NIR and X-ray flare are known.

\subsection{Initial electron Lorentz factor}\label{app1:gamma0}
\begin{figure}
    \centering
    \includegraphics[width=0.45\textwidth]{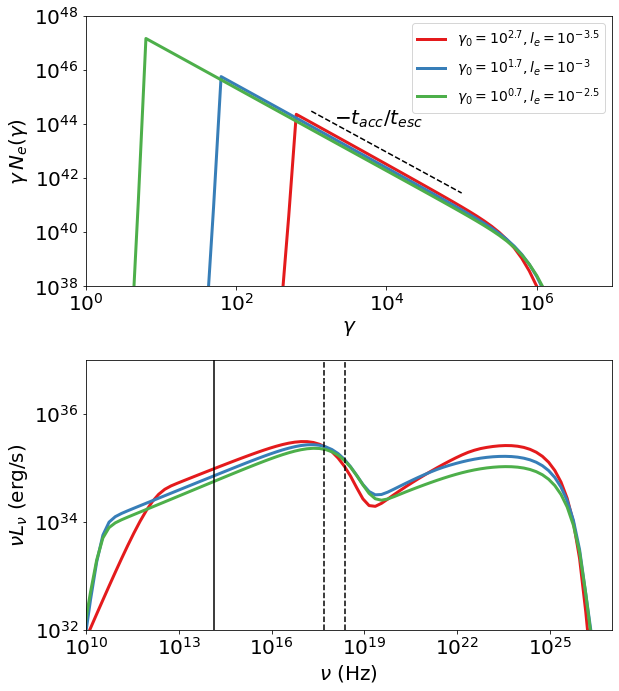}
    \caption{Effects of the minimum electron Lorentz factor on particle and photon spectra of a nonthermal flare. Top panel: Electron distribution $N_e(\gamma)$ multiplied by $\gamma$ for different values of the initial Lorentz factor $\gamma_0$ and electron compactness $\ell_e$ (see inset legend). All other parameters are the same as in Table~\ref{tab:param}. Bottom panel: SSC spectra produced by the electron distributions shown on the left panel. The snapshots shown in both panels are computed at $t=11 t_{acc}, 14.5 t_{acc}, 17 t_{acc}$ for $\gamma_0=10^{3.5}, 10^{2.7}$ and $10^{0.7}$ respectively.}
    \label{fig:sed-g0}
\end{figure}

To illustrate the effects of the Lorentz factor of electrons when injected in the acceleration region, we present in Fig.~\ref{fig:sed-g0} results for three values of $\gamma_0$ (see inset legend). We also adjust the injection compactness $\ell_e$ in each case so that the number of electrons at the cutoff of the distribution remains the same, thus producing the same level of X-ray emission, as shown on the bottom panel of the figure. Since particles are accelerated from different values of $\gamma_0$, the time needed for the cutoff Lorentz factor of the distribution to reach its saturation value increases with decreasing $\gamma_0$. Therefore, the choice of $\gamma_0$ affects the time delay between the onset of the NIR and X-ray flares. The broadband photon spectra are similar in all cases, with some differences appearing in the low-end of the synchrotron spectrum and the peak of the SSC spectrum. The synchrotron spectrum becomes self-absorbed around 10 GHz for lower values of $\gamma_0$ since the number of low-energy electrons increases, as shown in the top panel. 

\subsection{Energy-dependent timescales}\label{app1:tacc-gamma}
An important assumption made in our generic model is that both acceleration and escape timescales are constant 
(see Sect.~\ref{sec:model}), but more generally they can depend on the particle energy (or Lorentz factor). To exemplify the effects of the energy-dependent timescales, we adopt $t_{acc} = t_{a0}\gamma$,  $t_{esc} = t_{e0}\gamma$, and discuss the main differences with respect to our generic model. 

When $t_{acc}$ and $t_{esc}$ have the same dependence on $\gamma$, there is an analytical solution for the particle distribution \citep{KRM98}, which reads\footnote{We note a typo in Eq.~(A5) of \cite{KRM98}, which is missing a square root from the square brackets. We also note that  \cite{KRM98} use the symbol $\gamma_{\max}$ to denote the asymptotic value of the Lorentz factor, labeled as $\gamma_{sat}$ here.}
\begin{equation}\label{eq:Ne3}
    N_e(\gamma,t) = \frac{Q_0t_{a0} \gamma_{sat}^2}{(\gamma_{sat}^2-\gamma^2)} \left[\frac{\gamma}{\gamma_{0}}\sqrt{\frac{\gamma_{sat}^2-\gamma_0^2}{\gamma_{sat}^2-\gamma^2}}\right]^{-\frac{t_{a0}}{t_{e0}}}\Theta[t-\tau(\gamma)],
\end{equation}
where $\gamma_{sat}$ is the saturation Lorentz factor given by,
\begin{equation}
\gamma_{sat} = \left(b_s t_{a0}\right)^{-1/2} \propto B^{-1} t_{a0}^{-1/2}
\label{eq:gsat2}
\end{equation}
and $\tau$ is given by\footnote{The corresponding expression in  \cite{KRM98} (Eq.~A6) is missing the factor $\gamma_{sat}$ multiplying the logarithmic term.}
\begin{equation}
     \tau(\gamma) = \frac{t_{a0}\gamma_{sat}}{2}\ln\left[\frac{(\gamma_{sat}+\gamma)(\gamma_{sat}-\gamma_{0})}{(\gamma_{sat}-\gamma)(\gamma_{sat}+\gamma_0)}\right].
\end{equation}
For $\gamma \ll \gamma_{sat}$ the particle distribution is a power law with slope $s = t_{a0}/t_{e0}$, which is harder by 1 compared to the power-law spectrum obtained for constant timescales (see Eq.~\ref{eq:Ne}). The photon index of the emitted synchrotron spectrum is $s_{ph} = (1+t_{a0}/t_{e0})/2$, and differs accordingly by 1/2 compared to the value given by Eq.~\ref{eq:sph}. 

The condition imposed by the step function in Eq.~\ref{eq:Ne3} yields the time evolution of the high-energy cutoff of the particle distribution, 
\begin{equation}
\gamma_{\max}(t) = \gamma_{sat}\frac{\frac{\gamma_{sat}+\gamma_0}{\gamma_{sat}-\gamma_0} - e^{-\frac{2t}{ t_{a0}\gamma_{sat}}}}{\frac{\gamma_{sat}+\gamma_0}{\gamma_{sat}-\gamma_0} +  e^{-\frac{2t}{t_{a0}\gamma_{sat}}}} \approx \gamma_{sat} \tanh\left( \frac{t}{t_{a0}\gamma_{sat}}\right),
\label{eq:gmax2}
\end{equation}
where the approximation holds when $\gamma_{sat} \gg \gamma_0$.   
 
\begin{figure}
\centering 
\includegraphics[width = 0.45\textwidth]{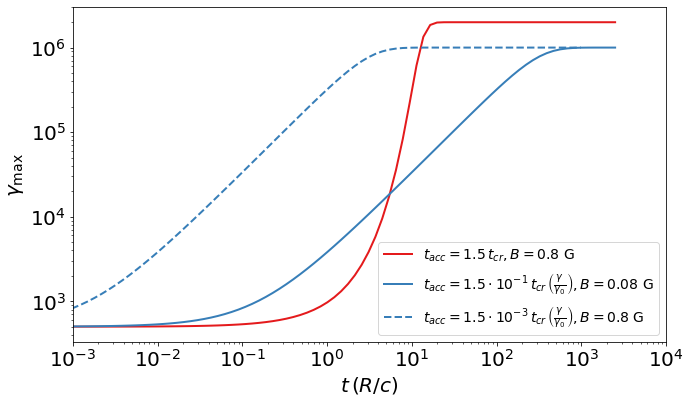}
\caption{Time evolution of the particle distribution high-energy cutoff computed for constant (red curve) and energy-dependent (blue curves) acceleration timescales according to Eqs.~\ref{eq:gmax} and \ref{eq:gmax2} respectively. For the parameters used, see inset legend.}
\label{fig:gmax}
\end{figure}

To better illustrate the differences with the high-energy cutoff evolution given by Eq.~\ref{eq:gmax}, we plot in Fig.~\ref{fig:gmax} $\gamma_{\max}(t)$  for a constant acceleration timescale (red curve) and an acceleration timescale scaling linearly with particle energy (blue curves). For the latter case, we show results for $t_{a0} = 0.15 \, t_{cr}/\gamma_0, B=0.08$~G (solid line) and $t_{a0} = 0.0015 \, t_{cr}/\gamma_0, B=0.8$~G (dashed line), while $t_{a0}/t_{e0}=1.5$ (where $t_{cr}=R/c$).  All other parameters are the same as in Table~\ref{tab:param}. For the selected parameter values, the saturation Lorentz factor is similar in all three cases, but the gradient of the $\gamma_{\max}(t)$ curve is smaller when $t_{acc} \propto \gamma$. By appropriately choosing $t_{a0}$ and $B$ one controls $\gamma_{sat}$ and the time needed to reach this value. These choices then map to the observed maximal synchrotron photon energy, $\epsilon \propto B \gamma_{sat}^2$, and the rise time of the light curve at that energy. 
  
Using Eq.~\ref{eq:gsat2} we can estimate the magnetic field strength needed to radiate synchrotron photons at a known photon energy $\epsilon_X$ by electrons with Lorentz factor $\gamma_{sat}$, 
\begin{eqnarray}
B & = &  t_{a0}^{-1} \left(\frac{m_e c^2}{\epsilon_X}\right)  \left(\frac{6 \pi m_e c}{\sigma_T B_{cr}} \right) \nonumber  \\
& \simeq & 5~{\rm mG} \,  \left(\frac{(2s_{ph} -1)}{1.5}\right)^{-1} \left(\frac{R}{9.5 R_S} \right)^{-1}\left(\frac{\epsilon_X}{10~\rm keV}\right)^{-1}.
\label{eq:B2}
\end{eqnarray}
The numerical value above was obtained for $t_{a0}/t_{e0}=1.5$, $t_{a0} = 0.15~t_{cr}/\gamma_0 \simeq 0.08$~s, $\gamma_0=10^{2.7}$, and $R=9.5 R_S$. Unless the acceleration timescale is much faster than the photon crossing timescale (i.e., $t_{a0}\gamma_0 \ll t_{cr}$; see also blue dashed line in Fig.~\ref{fig:gmax}), X-ray emitting electrons have to gyrate in $\sim$mG magnetic fields. Such values are disfavored by modeling of the sub-mm emission of \sgras with synchrotron radiation of quasi-thermal electrons, which yields magnetic field strengths of $\sim 10$~G in the inner accretion flow \citep[see e.g.][]{Genzel2010, EHT_2022}. Moreover, the energy density of nonthermal electrons in the blob, 
\begin{equation}
u_e = \frac{3 m_e c^2}{4 \pi R^3} \int_{\gamma_0}^{\gamma_{sat}} d\gamma \, \gamma N_{e}(\gamma) \approx \frac{3 m_e c^2}{4 \pi R^3} \frac{Q_0 t_{a0} \gamma_{sat}^2}{2-s}\left( \frac{\gamma_{sat}}{\gamma_0}\right)^{-s}
\end{equation}
is many orders of magnitude larger than the magnetic energy density for mG fields, $u_B = B^2/(8\pi)$, resulting in SSC-dominated photon spectra and electron cooling (see also Appendix~\ref{app1:radius}). In conclusion, a flare model for \sgras with $t_{acc}=t_{a0}\gamma$ and $t_{esc}=t_{e0} \gamma$ would require $t_{a0}, t_{e0} \ll t_{cr}$.

\subsection{Doppler boosting}\label{app1:doppler}
Astrometric observations performed by the \gravity\ interferometer during NIR flares have revealed that the NIR emitting region moves along loop trajectories with a projected radius on the plane of the sky of of $\sim (6-10)\cdot R_g$ \citep{gravity2018flares, GRAVITY2020}. Here, $R_g = R_S/2$ is the gravitational radius of the black hole. Some interpretations of these observations include a compact region in the accretion flow (hot spot) moving on a circular orbit at a radius $\sim 9 R_g$ around the black hole \citep{GRAVITY2020} or a compact region (blob/plasmoid) orbiting the funnel region of the accreting black-hole system \citep{Ball_2021, 2024MNRAS.531.3136L}. Such structures, which are the result of magnetic reconnection, have been identified in general-relativistic magnetohydrodynamic (GRMHD) simulations \citep{Nathanail_2022, Ripperda_2022, 2024MNRAS.531.3136L}. 

To examine the effects of Doppler boosting on the light curves expected in our baseline model (see Fig.~\ref{fig:lc-decay}) arising from the relative motion of the emitting region and a distant observer, we adopt the scenario described in \cite{Ball_2021}. We assume that the flaring region is a blob that moves on a conical spiral trajectory.  The equations of motion for the blob (in spherical coordinates)  are:
\begin{eqnarray}
 r(\tau) & = &  r_0 + \upsilon_r \tau \\ 
 \theta(\tau) & = &  \theta_0 \\ 
 \dot{\phi}(\tau) r^2(\tau) & = &   \dot{\phi}_0 r^2_0
\end{eqnarray}
where $\tau = c t / R_g$, $r= R/R_g$, and $\upsilon_r$ is in units of $c$. The initial conditions are indicated by the subscript $0$. The azimuthal velocity of the blob is given by $\upsilon_\phi(\tau) = r(\tau) \sin(\theta_0)\dot{\phi}(\tau)$. The Doppler factor of the blob is then defined as:

\begin{equation}
\delta(\tau) = \frac{1}{\Gamma_b(\tau) (1- \vec{\upsilon}_b \cdot \vec{n}_{obs})},
\end{equation}
where $\Gamma_b = (1-\upsilon_b^2(\tau))^{-1/2}$ , $\upsilon_b =(\upsilon_r^2 + \upsilon_\phi^2)^{1/2}$ is the blob velocity (in units of $c$), and $\vec{n}_{obs}$ is a unit vector indicating the observer's location. The photon frequency and integrated luminosities (between $\nu_1$ and $\nu_2$) as measured by a distant observer are then given by   $\nu_{obs} = \delta \nu$, and $L_{obs}= \delta^4 \int_{\nu_1}^{\nu_2} {\rm d} \nu L_{\nu}$, while the time interval between the arrival of two photons in the observer's frame is $d\tau_{obs} = d\tau / \delta(\tau)$.

We adopt the initial conditions presented in Table~1 of \cite{Ball_2021}: $r_0=36, \upsilon_r = 0.01, \upsilon_{\phi_0} = 0.41, \phi_0 = 200^{\rm o}, \theta_0 = 15^{\rm o}, \theta_{\rm obs}=168^{\rm o}$ and $\phi_{obs}=90^{\rm o}$. We vary $\phi_0, \upsilon_{\phi_0}$ and $v_r$ to illustrate their impact on the light curve shape. Throughout our calculations we assume that the conditions in the blob, such as $B$ and $R$ do not change with time, and are the same as those in listed in Table~\ref{tab:param}. We do not account for gravitational redshift and lensing. Our results are presented in Figs.~\ref{fig:lc-Xray-boost}-\ref{fig:lc-NIR-boost-vr}. Generally, Doppler boosting (or deboosting) of the emitted radiation by the blob produces multi-peaked light curves, thus increasing the complexity of the flare profiles in the baseline model.  

\begin{figure}
\centering
\includegraphics[width=0.45\textwidth]{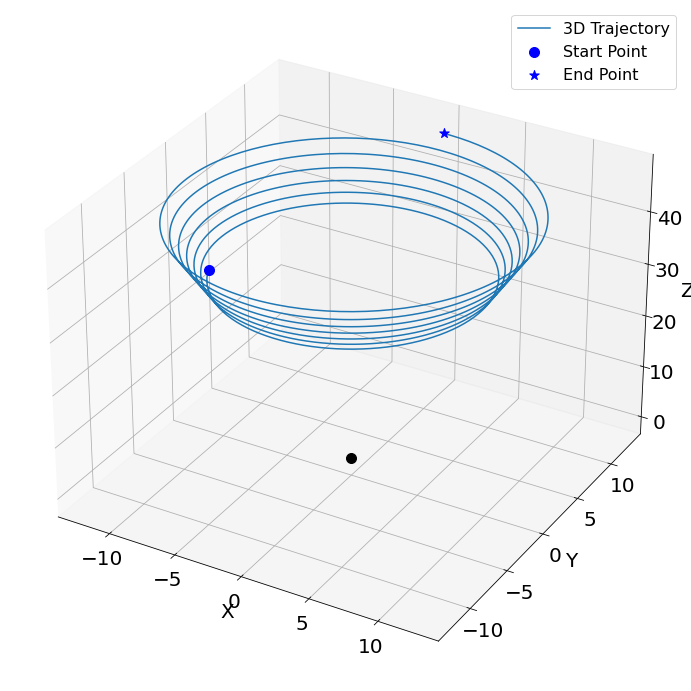}
\caption{Blob trajectory computed for a period of $1300~R_g/c \approx 68 R/c$ where the blob radius is $R\approx 9.5 R_S$. For the initial conditions, see text.}
\label{fig:3d}
\end{figure}

\begin{figure}
\centering
\includegraphics[width = 0.47\textwidth]{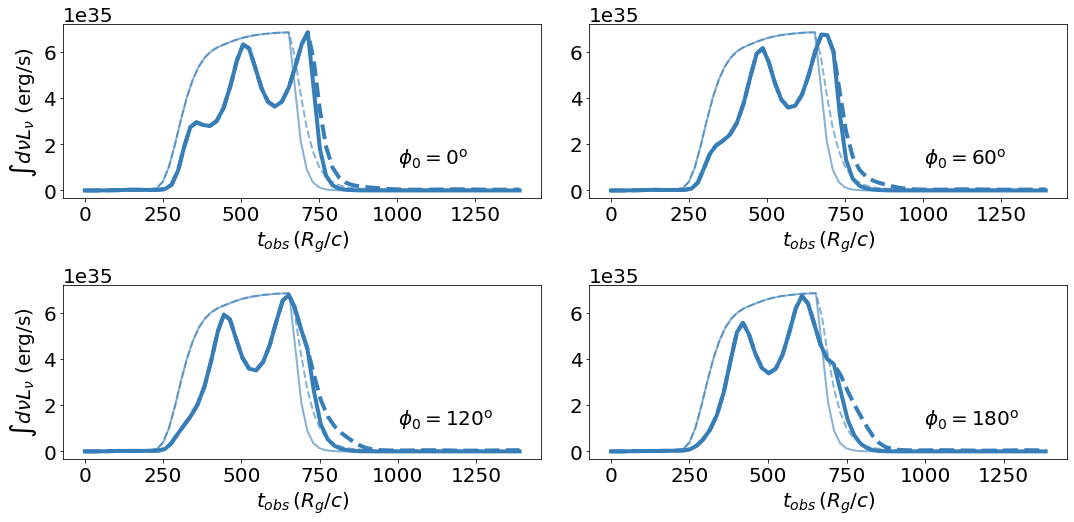}
\caption{X-ray (2-10 keV) light curves (thick lines) computed for a blob moving on a conical spiral trajectory, starting from different initial azimuth angles (see inset legends). The observer is located at $\theta_{obs}=168^{\rm o}$ in the $y-z$ plane -- see also Fig.~\ref{fig:3d}. Thin lines show the light curves without the Doppler boosting effects, as in Fig.~\ref{fig:lc-decay}.}
\label{fig:lc-Xray-boost}
\end{figure}

\begin{figure}
\centering
\includegraphics[width = 0.47\textwidth]{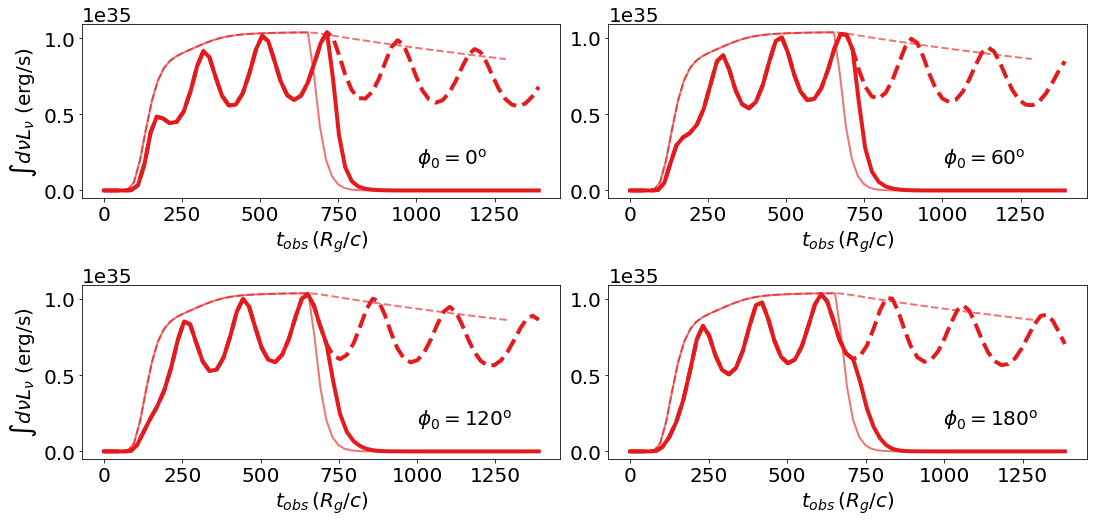}
\caption{Same as in Fig.~\ref{fig:lc-Xray-boost} but for the NIR band (2.2 $\mu m$).}
\label{fig:lc-NIR-boost}
\end{figure}

\begin{figure}
\centering
\includegraphics[width = 0.47\textwidth]{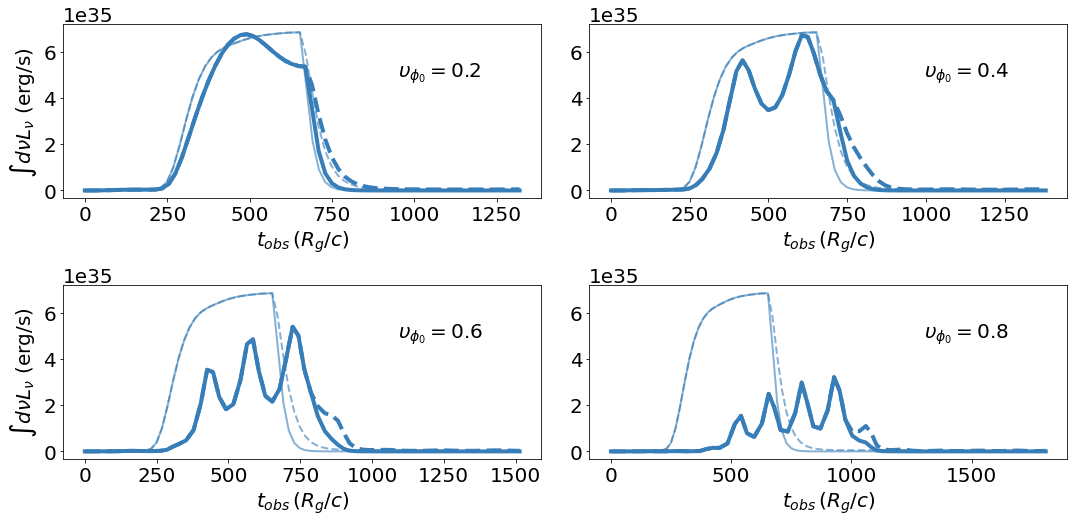}
\caption{Same as in Fig.~\ref{fig:lc-Xray-boost} but for different initial azimuthal velocities.}
\label{fig:lc-Xray-boost-vphi}
\end{figure}

\begin{figure}
\centering
\includegraphics[width = 0.47\textwidth]{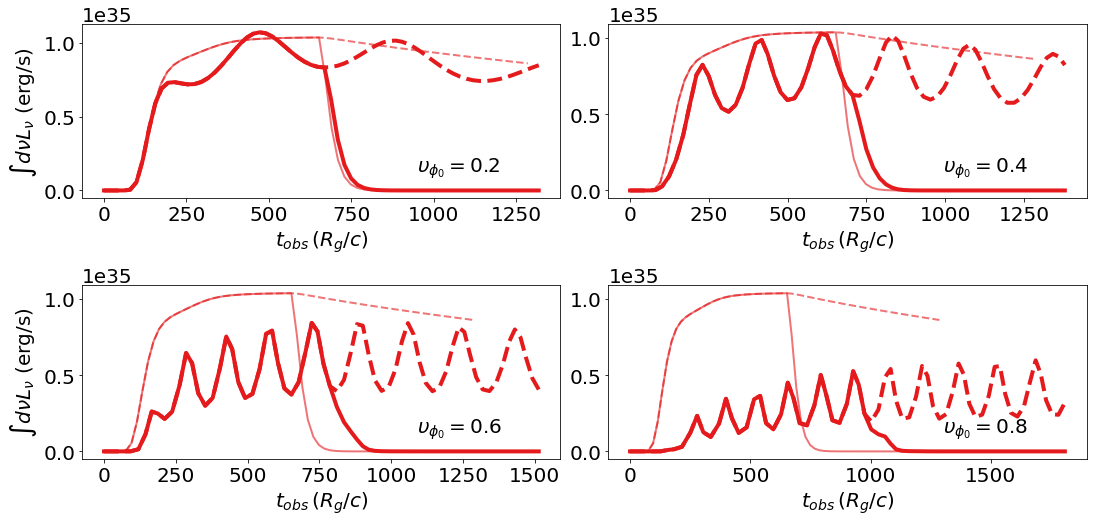}
\caption{Same as in Fig.~\ref{fig:lc-NIR-boost} but for different initial azimuthal velocities.}
\label{fig:lc-NIR-boost-vphi}
\end{figure}

\begin{figure}
\centering
\includegraphics[width = 0.47\textwidth]{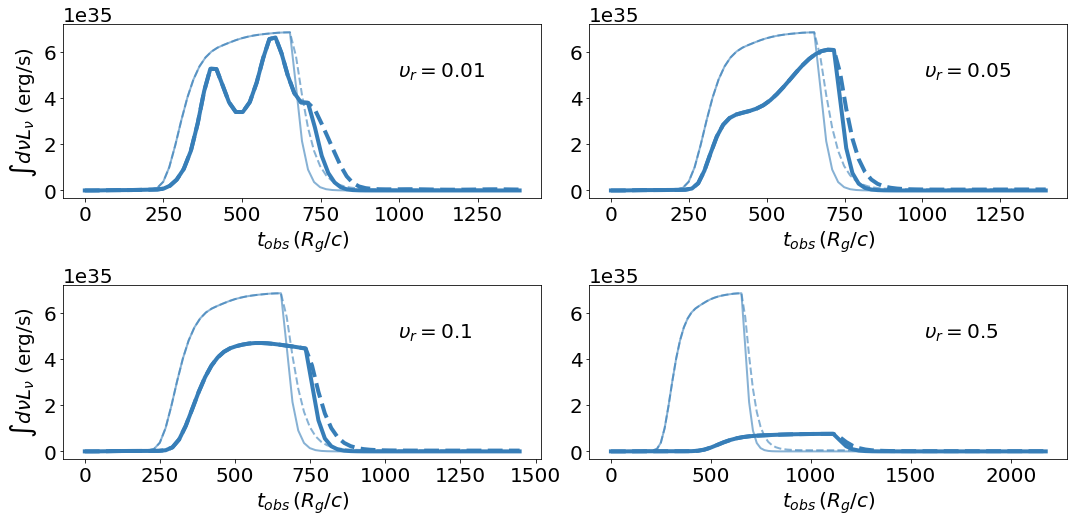}
\caption{Same as in Fig.~\ref{fig:lc-Xray-boost} but for different radial blob velocities.}
\label{fig:lc-Xray-boost-vr}
\end{figure}

\begin{figure}
\centering
\includegraphics[width = 0.47\textwidth]{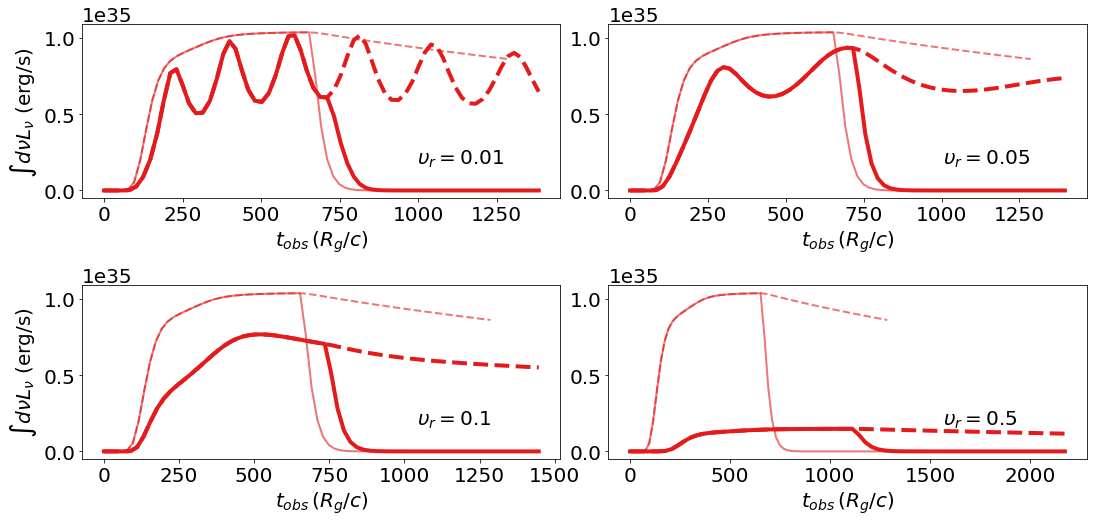}
\caption{Same as in Fig.~\ref{fig:lc-NIR-boost} but for different radial blob velocities.}
\label{fig:lc-NIR-boost-vr}
\end{figure}

\end{document}